\documentclass[12pt,tightenlines,eqsecnum,floats,aps,amsmath,amssymb,nofootinbib,prd,superscriptaddress,showpacs]{revtex4}

\usepackage{setspace}
\usepackage{amsmath,amssymb,amsfonts,amsthm}
\usepackage{graphicx}
\usepackage{enumerate}

\makeatletter
\newcommand*{\simboloG}[1]{%
  \vphantom{\sum}
  \smash{%
    \mathchoice{%
      \raisebox{-.3\height}{\Huge$\m@th\displaystyle#1$}%
      }{%
      \raisebox{-.2\height}{\LARGE$\m@th#1$}%
      }{%
      \raisebox{-.2\height}{\LARGE$\m@th#1$}%
      }{%
      \raisebox{-.2\height}{\LARGE$\m@th#1$}%
      }%
    }}
\newcommand{\BigTimes}{\mathop{\simboloG{\times}}}
\makeatother

\newcommand{\GT}{G_{{\scriptscriptstyle \mathbb{T}^2}}}
\newcommand{\GDL}{G_{{\scriptscriptstyle \mathbb{S}^2}}^{{\scriptscriptstyle \mathrm{DLM}}}}
\newcommand{\GGM}{G_{{\scriptscriptstyle \mathbb{S}^2}}^{{\scriptscriptstyle \mathrm{GM}}}}

\begin{document}

\title{Generating functions for black hole entropy in Loop Quantum Gravity}

\author{J. Fernando \surname{Barbero G.}}
\email[]{fbarbero@iem.cfmac.csic.es} \affiliation{Instituto de
Estructura de la Materia, CSIC, Serrano 123, 28006 Madrid, Spain}

\author{Eduardo J. \surname{S. Villase\~nor}}
\email[]{ejsanche@math.uc3m.es} \affiliation{Instituto Gregorio Mill\'an, Grupo de Modelizaci\'on
y Simulaci\'on Num\'erica, Universidad Carlos III de Madrid, Avda.
de la Universidad 30, 28911 Legan\'es, Spain} \affiliation{Instituto
de Estructura de la Materia, CSIC, Serrano 123, 28006 Madrid, Spain}

\date{April 30, 2008}

\begin{abstract}
We introduce, in a systematic way, a set of generating functions that solve \textit{all} the different combinatorial problems that crop up in the study of black hole entropy in Loop Quantum Gravity. Specifically we give generating functions for: The different sources of degeneracy related to the spectrum of the area operator, the solutions to the projection constraint, and the black hole degeneracy spectrum. Our methods are capable of handling the different countings proposed and discussed in the literature. The generating functions presented here provide the appropriate starting point to extend  the results already obtained for microscopic black holes  to the macroscopic regime --in particular those concerning the area law and the appearance of an effectively equidistant area spectrum.
\end{abstract}


\maketitle

\section{Introduction} The study of the black hole degeneracy spectrum in Loop Quantum Gravity (LQG) has provided important support for the formalism. The confirmation that the expected behavior for the entropy as a function of area is obtained is one of the main \textit{physical} achievements  claimed in this framework \cite{abk}. In addition to the early successes in this respect there has been an important series of results in the recent past related to this problem. In particular, the studies carried out in \cite{val1,val2} have unearthed a very rich and unexpected behavior of the black hole degeneracy spectrum predicted by LQG. For microscopic black holes these papers show that, in addition to the exponential growth compatible with the Bekenstein-Hawking area law, an effective equi-spacing of the spectrum (with a period approximately proportional to $\log 3$) is present. If this feature survives in the macroscopic limit it would be a very interesting consequence of LQG because, despite the unevenness of the spectrum of the area operator, an emergent, effective, regular spacing would be predicted. This can be seen as an additional consistency check for the formalism because such a behavior is expected on general grounds \cite{BM}.

In order to extend the existing microscopic results to macroscopic areas one has to find a way to reach this asymptotic limit without spoiling the content of the theory with un-controllable approximations. This is very much in the spirit of Mathematical Statistical Mechanics and Combinatorics. As it is usually done in that framework, the process of reaching the asymptotic (thermodynamical) limit requires a number of steps. The first one is casting the problem at hand --in this case the counting of the relevant black hole micro-states-- in such a way that the intimate mathematical nature of the model is captured. This has already been done in \cite{prlnos} for the black hole entropy problem in LQG. The success in this first step can be judged by trying to carry out the second one: Obtaining suitable generating functions for the combinatorial problems involved in the counting of states. This is the purpose of this paper. A third --and final-- problem, that has to be tackled immediately after the one considered here, is to get appropriate asymptotic expansions capturing the macroscopic behavior of black hole entropy as predicted by LQG. This may well be the hardest step due to its analytic nature. Almost certainly it  will require mathematical tools different from the number-theoretical and combinatorial methods used to complete the first two parts of the program described above.

The paper is organized as follows. After this introduction we review in section \ref{area} the algorithm proposed in \cite{prlnos} to compute the black hole degeneracy. The notation used in the paper will closely follow that of \cite{prlnos}. Section \ref{diophant} is devoted to the obtention of the generating functions counting the number of solutions to the linear diophantine equations needed to describe the degeneracy of the area operator. Section \ref{BH} deals with the generating functions giving the full black hole degeneracy spectrum for the different versions of the projection constraint that appear in the literature. Finally we end with some conclusions and comments in section \ref{conclusions}.

All the results presented here refer to the isolated horizons that are used to model black holes in LQG. When we talk about black hole properties in the following we refer, in fact, to the isolated horizons representing them.

\section{Characterization of the area spectrum}\label{area}

The black hole area in LQG is given by eigenvalues
$A$ of the area operator of the form
\begin{equation}
A=8\pi\gamma
\ell_P^2\sum_{I=1}^N\sqrt{j_I(j_I+1)}\,,\label{area_eigenvalue}
\end{equation}
where $\gamma$ is the Immirzi parameter and $\ell_P$ is the Planck length. Here the labels $j_I\in\mathbb{N}/2$ are half-integers associated to the edges of a certain spin network state. They pierce the isolated horizon representing the black hole at a finite set
of $N$ points called punctures \cite{abk}. In the following we will choose units such that $4\pi\gamma\ell_P^2=1$. Horizon quantum states are further characterized by an extra label $m_I$ that can be interpreted as a spin component. Depending on the horizon topology these labels are restricted to satisfy certain constraints that we will discuss later.

The real numbers belonging to the spectrum of the area operator have been characterized in \cite{prlnos}. An obvious, but important, comment is that these numbers must be linear combinations of square roots of square-free numbers (SRSFN) $p_i$ with non-negative integer coefficients $q_i$. In order to check if a number $a=\sum_{i=1}^rq_i\sqrt{p_i}$ belongs to the area spectrum there must exist $j_I:=k_I/2$, $k_I\in \mathbb{N}$, such that
\begin{equation}
\sum_{I=1}^N\sqrt{(k_I+1)^2-1}=\sum_{k=1}^{k_{\mathrm{max}}}n_k\sqrt{(k+1)^2-1}=\sum_{i=1}^rq_i\sqrt{p_i}=a.
\label{areaequation}
\end{equation}
Here the $n_k$  denote
the number of punctures corresponding to edges carrying spin $k/2$; hence the sum $n_1+\cdots n_{k_{\mathrm{max}}}=N$ is just the total number of punctures.
Notice that we can always write $\sqrt{(k+1)^2-1}$ as the product of an integer times the square root
of a square-free positive integer number (SRSFN) by using its prime
factor decomposition. Equation (\ref{areaequation}) is solved in two steps: First we must identify the allowed spins $k/2$ such that $\sqrt{(k+1)^2-1}$ is an integer multiple of some $\sqrt{p_i}$, and then determine the value of $n_k$ that tells us how many times each of them appears. In order to deal with the first problem we must solve the Pell equations associated to each of the SRSFN's in the r.h.s. of (\ref{areaequation}), i.e.
\begin{equation}
\label{pell} \sqrt{(k+1)^2-1}=y \sqrt{p_i}\,\Leftrightarrow (k+1)^2-p_iy^2=1, 
\end{equation}
with $y\in\mathbb{N}$. We will label the solutions as $\{(k_m^i,y_m^i)\,:\,m\in\mathbb{N}\}$, where the index $i$ refers to the square-free numbers in each of the Pell equations (see, for instance, \cite{Burton} for details on the Pell equation).  Once these numbers are known the $n_k$ can be found by solving the system of $r$-uncoupled \cite{prlnos}, linear, diophantine equations
\begin{equation}
\sum_{m=1}^\infty y_m^i n_{k_m^i}=q_i,\quad i=1,\ldots, r.
\label{diof}
\end{equation}
Notice that, once the $q_i$ are fixed, only a finite number of spins $k^i_m/2$, $m=1,\ldots,M_i$, come into play in the equations (\ref{diof}).

It may happen that some of these equations admit no solutions.
In this case $\sum_{i=1}^r q_i \sqrt{p_i}$ does not
belong to the area spectrum. On the other hand, if
they do admit solutions, the $\sum_{i=1}^r q_i
\sqrt{p_i}$ belong to the spectrum of the area operator, the numbers
$k_m^i$ tell us the spins involved, and the  $n_{k_m^i}$ count the
number of times that the edges labeled by the spin $k_m^i/2$ pierce
the horizon.

Let us denote by $\mathcal{S}^i_{q_i}$, $i=1,\ldots, r$, the set built from the solutions to the $i$-th diophantine equation appearing in (\ref{diof}) as $$\mathcal{S}^i_{q_i}=\big\{s_i=\{(k_m^i,n_{k_m^i})\}_{m=1}^{M_i}\,:\,\sum_{m=1}^{M_i} y_m^i n_{k_m^i}=q_i \big\}.$$
The elements in these sets are combined in the Cartesian product   $\mathcal{S}_a=\BigTimes_{i=1}^r \mathcal{S}^i_{q_i}$ to give all the solutions to the system (\ref{diof}). The set $\mathcal{S}_a$ contains all the \textit{spin configurations} $s=(s_1,s_2,\ldots,s_r)\in \mathcal{S}_a$  defined by the area $a=\sum_{i=1}^r q_i\sqrt{p_i}$.

Once we have these configurations the black hole degeneracy spectrum is obtained as
\begin{equation}
e^{S(a)}:=\sum_{s\in\mathcal{S}_a}\frac{(\sum_{i=1}^r\sum_{(k,n_k)\in s_i} n_k)!}{\prod_{i=1}^r\prod_{(k,n_k)\in s_i} n_k!}P(s),
\label{deg}
\end{equation}
where the sum $\sum_{(k,n_k)\in s_i}$ and product $\prod_{(k,n_k)\in s_i}$ are extended to the elements $(k,n_k)$ of the $i$-th component $s_i$ of the spin configuration $s$. The factor $P(s)=P(s_1,\ldots,s_r)$ is introduced to take into account the projection constraint. The different choices for $P$ will be discussed in the following sections.

\section{Diophantine equations}\label{diophant}

The purpose of this section is twofold. We will first introduce a generating function giving the number of solutions for a collection of uncoupled diophantine equations of the form given by (\ref{diof}). Afterwards we will modify this generating function in order to get the reordering degenerations given by the sum of multinomial coefficients in (\ref{deg}) obtained by taking $P(s)=1$ for every configuration.

The generating functions that we will discuss in the following are written in terms of variables that we will denote as $x_i$ with $i$ in one-to-one correspondence with the square-free numbers $p_i$. They involve numerical constants that are obtained from the solutions to the Pell equations for each $p_i$; in particular the numbers $k_m^i$ and $y_m^i$ introduced in the previous section. For each of the diophantine equations given by (\ref{diof}) the generating function counting the number of its solutions can be found in any text book on Discrete Mathematics or Combinatorics (see, for example, \cite{DM}). It has the following simple form
\begin{equation}
G_i^\textrm{\#\,sol}(x_i)=\prod_{m=1}^\infty\frac{1}{(1-x_i^{y^i_m})}.
\label{Gi}
\end{equation}
The coefficient of $x_i^{q_i}$ in the Taylor expansion of (\ref{Gi}) around $x_i=0$ gives the number of non-negative solutions to the corresponding diophantine equation (\ref{diof}). Notice that, although we are writing an infinite product, in every case we only need a finite number of $y^i_m$ (those smaller or equal to $q_i$) in order to determine the required coefficient.

The total number of solutions for a system of such uncoupled diophantine equations is just given by the product of the individual generating functions
\begin{equation}
G^\textrm{\#\,sol}(x_1,x_2,\ldots)=\prod_{i=1}^\infty G_i^\textrm{\#\,sol}(x_i)=\prod_{i=1}^\infty\prod_{m=1}^\infty\frac{1}{(1-x_i^{y^i_m})}.
\label{Gsol}
\end{equation}
Notice again that for a fixed value of the area only a finite number of square-free $p_i$ will be involved and, hence, the infinite product $\prod_{i=1}^\infty$ is, in fact, finite.

The generating function $G^\textrm{\#\,sol}$ just computes $ \sum_{s\in\mathcal{S}_a} \!1$ for each allowed value of the area (and it gives zero if $\mathcal{S}_a=\emptyset$). The coefficient of the term $x_1^{q_1}\cdots x_r^{q_r}$ in the Taylor expansion of $G^\textrm{\#\,sol}$ is the number of solutions to the system of simultaneous  diophantine equations (\ref{diof}) and, hence, it coincides with $ \sum_{s\in\mathcal{S}_a} \!1$. Now we want to modify (\ref{Gsol}) in such a way that we obtain a generating function for the sum of multinomial numbers
\begin{equation}
\sum_{s\in\mathcal{S}_a}\frac{(\sum_{i=1}^r\sum_{(k,n_k)\in s_i} n_k)!}{\prod_{i=1}^r\prod_{(k,n_k)\in s_i} n_k!}\,.\label{factorial}
\end{equation}
A simple way to do it is following a two step approach: First we modify (\ref{Gi}) to introduce the product of factorials in the denominator of (\ref{factorial}) in front of each term of its Taylor expansion. This can easily be done by considering
$$
\exp\Big(\sum_{i=1}^\infty\sum_{m=1}^\infty x_i^{y^i_m}\Big)\,.
$$
We still have to introduce the factorial appearing in the numerator of (\ref{factorial}). This can be done by manipulating  the previous expression in the following formal way. Let us take
$$
G^{\mathrm{aux}}(\omega;x_1,x_2,\ldots)=\int_0^\infty  e^{-\lambda}\, \exp\Big(\lambda \omega \sum_{i=1}^\infty\sum_{m=1}^\infty x_i^{y^i_m}\Big)\,\mathrm{d}\lambda
$$
and consider $G^{\mathrm{aux}}(\omega;x_1,x_2,\ldots)$ for $\omega=1$. It can be readily seen that
\begin{eqnarray*}
G^{\textrm{d}}(x_1,x_2,\dots)=G^{\mathrm{aux}}(1;x_1,x_2,\ldots)=\left(\displaystyle 1-\sum_{i=1}^\infty\sum_{m=1}^\infty x_i^{y^i_m}\right)^{-1}
\end{eqnarray*}
has the required form. This is a consequence of the following simple formal argument: If $f(x)=\sum_{n=0}^\infty a_n x^n$ then the function $g(\omega)$, whose Taylor coefficients are $n! a_n$, is given in terms of $f$ by
$$
g(\omega)=\int_0^\infty e^{-\lambda}f(\lambda \omega)\mathrm{d}\lambda= \int_0^\infty e^{-\lambda}\left(\sum_{n=0}^\infty a_n \lambda^n\omega^n\right)\mathrm{d}\lambda=\sum_{n=0}^\infty  n! a_n\omega^n\,.
$$

\section{Generating function for the black hole degeneracy spectrum}\label{BH}

Let us consider now other choices for $P$ in (\ref{deg}). Some of them have a direct physical meaning whereas others allow us to discuss other possible projection constraints similar in form to the standard ones.

\subsection{Toroidal black holes}

This case corresponds to considering
$$
P_{{\scriptscriptstyle \mathbb{T}^2}}(s)=\prod_{i=1}^r\prod_{(k,n_k)\in s_i} (k+1)^{n_k}.
$$
This choice describes a situation in which the third spin components $m_I$ are unconstrained and can take any of the $k_I+1$ possible values independently of each other. In the literature this is expressed by saying that no projection constraint is involved \cite{val1} and it can be shown that it describes \textit{toroidal} black holes \cite{Kloster:2007cb}. The relevant generating function is
\begin{equation}
\GT(x_1,x_2,\dots)=\left(\displaystyle 1-\sum_{i=1}^\infty\sum_{m=1}^\infty (k^i_m +1) x_i^{y^i_m}\right)^{-1}\,.
\label{GT}
\end{equation}
The coefficient of term $x_1^{q_1}\cdots x_r^{q_r}$ in the Taylor expansion of the previous expression is the total degeneracy (\ref{deg}) of a toroidal horizon with area given by $q_1\sqrt{p_1}+\cdots+q_r\sqrt{p_r}$. The function $\GT$ can be obtained as before in two steps. Consider first
$$
\exp\Big(\sum_{i=1}^\infty\sum_{m=1}^\infty (k^i_m+1)x_i^{y^i_m}\Big),
$$
that produces the required inverse factorial terms and also the product $\prod_{i=1}^r\prod_{(k,n_k)\in s_i} (k+1)^{n_k}$, and then introduce the factorial term in the numerator of (\ref{deg}) by using the same formal trick described at the end of section \ref{diophant}. As in previous instances the formal infinite products and sums  in (\ref{GT}) are, in practice, finite because only a finite number of square-free integers are involved for a fixed area value. This means that the generating function can be considered, for concrete computations, as a rational function with a finite number of variables.

\subsection{Spherical black holes}

In the case where we have spherical symmetry the so called \textit{projection constraint}
\begin{equation}
\sum_{I=1}^Nm_I=0\label{proy_const}
\end{equation}
must be satisfied by the spin components $m_I$. The accepted view in LQG \cite{DL}, that we will refer to as the DLM counting, is that each $m_I$ is further constrained to satisfy $m_I\in\{- k_I/2,k_I/2\}$. There are other proposals in the literature, in particular the GM counting of \cite{GM}, where a different prescription $m_I\in\{-k_I/2,-k_I/2+1,\ldots,k_I/2-1,k_I/2\}$ is suggested. From a purely combinatorial point of view both can be treated in a very similar way so in the following we will give generating functions for both approaches.

The new ingredient that we need in order to take into account the projection constraint is a suitable way to count  the number of solutions to (\ref{proy_const}). This can be done in a straightforward way. For the standard DLM counting, once the values of $k_I$ at the punctures are given, the number of solutions to the projection constraint is the constant term in the Laurent expansion of
$$
\prod_{I=1}^N(z^{k_I}+z^{-k_I}),
$$
whereas for the GM counting the number of solutions to the projection constraint is the constant term in the Laurent expansion of
$$
\prod_{I=1}^N\sum_{\alpha=0}^{k_I}z^{k_I-2\alpha}=\prod_{I=1}^N\frac{z^{k_I+1}-z^{-k_I-1}}{z-z^{-1}}.
$$
The generating function in these cases can be easily obtained from the toroidal one (\ref{GT}) by taking now
\begin{eqnarray*}
P_{{\scriptscriptstyle \mathbb{S}^2}}^{{\scriptscriptstyle \textrm{DLM}}}(s,z)&=&\prod_{i=1}^r\prod_{(k,n_k)\in s_i} (z^{k}+z^{-k})^{n_k}\,,\\
P_{{\scriptscriptstyle \mathbb{S}^2}}^{{\scriptscriptstyle \textrm{GM}}}(s,z)&=&
\prod_{i=1}^r\prod_{(k,n_k)\in s_i}
\Big(\sum_{\alpha=0}^{k}z^{k-2\alpha}\Big)^{n_k}=
\prod_{i=1}^r\prod_{(k,n_k)\in s_i} \Big(\frac{z^{k+1}-z^{-k-1}}{z-z^{-1}}\Big)^{n_k}\,.
\end{eqnarray*}
In view of the structure of the $P_{{\scriptscriptstyle \mathbb{S}^2}}^{{\scriptscriptstyle \textrm{DLM}}}(s,z)$ and $P_{{\scriptscriptstyle \mathbb{S}^2}}^{{\scriptscriptstyle \textrm{GM}}}(s,z)$ terms we can get the desired generating function by substituting the $(k_m^i+1)$ term in (\ref{GT}) for $(z^{k_m^i}+z^{-k_m^{i}})$ or $\Big(\sum_{\alpha=0}^{k^i_m} z^{k^i_m-2\alpha}\Big)$ respectively. This way we obtain
\begin{eqnarray}
\GDL(z,x_1,x_2,\dots)&=&\left(\displaystyle 1-\sum_{i=1}^\infty\sum_{m=1}^\infty (z^{k^i_m} +z^{-k^i_m}) x_i^{y^i_m}\right)^{-1}\,,\label{GDL}\\
\GGM(z,x_1,x_2,\dots)&=&\left(\displaystyle 1-\sum_{i=1}^\infty\sum_{m=1}^\infty \Big(\sum_{\alpha=0}^{k^i_m} z^{k^i_m-2\alpha}\Big) x_i^{y^i_m}\right)^{-1}\,.\label{GGM}
\end{eqnarray}
These functions have an extra auxiliary argument $z$ that is not present in (\ref{GT}). The coefficient of the term $z^nx_1^{q_1}\cdots x_r^{q_r}$ tells us the value of the sum (\ref{deg}) with a projection constraint given by the condition
$$
\sum_I m_I=n.
$$
Notice that, at variance with the cases discussed in the previous sections, the exponents of $z$ can be negative and, hence, the expansions that we have to use are Laurent series in $z$. The choice $n=0$ corresponds to the spherical black holes.

\section{Conclusions and comments}\label{conclusions}

We have given a collection of generating functions for a series of combinatorial problems related to the description of the black hole degeneracy spectrum in Loop Quantum Gravity. The coefficients of their power series expansions give us the \textit{exact} solution to the counting problems that we want to solve. In particular, the generating functions (\ref{GDL}) and (\ref{GGM}) give us the spherical black hole degeneracy spectrum for the different countings considered here, whereas (\ref{GT}) gives the one corresponding to the toroidal case. For horizons of higher genus it is expected that similar formulas hold \cite{Kloster:2007cb}.

We want to end with some comments. The first is that, despite the apparent infinite number of terms involved in the different sums and products appearing in the paper, for a given value of area only finite numbers of variables and terms are needed. It is only the fact that the diophantine equations that we need to solve have an effective number of variables that depends on the area, that forces us to introduce a formally infinite number of them. 

To convince the reader of the power of this generating function techniques we give here a concrete numerical example: For an area $a=40\sqrt{2}+40\sqrt{3}$ the number of possible configurations can be computed by the considering the generating function
$$
G^{\#\,\mathrm{sol}}(x_1,x_2)=\frac{1}{(1-x_1^2)(1-x_1^{12})(1-x_2)(1-x_2^4)(1-x_2^{15})}
$$
and extracting the coefficient of the term $x_1^{40}x_2^{40}$ which has a value of 84. The total degeneracy (in the DLM counting) is obtained by taking the generating function
\begin{eqnarray*}
&&\GDL(z,x_1,x_2)=\\
&&=\frac{1}{1-(z^2+z^{-2})x_1^2-(z^{16}+z^{-16})x_1^{12}-(z+z^{-1})x_2-(z^{6}+z^{-6})x_2^{4}-
(z^{25}+z^{-25})x_2^{15}}\,.
\end{eqnarray*}
The value of the black hole degeneracy $e^{S(40\sqrt{2}+40\sqrt{3})}$ is given by the coefficient of the $z^0x_1^{40}x_2^{40}$ in the power series expansion of $\GDL(z,x_1,x_2)$. This is 
$$
e^{S(40\sqrt{2}+40\sqrt{3})}=991809938488860909241077458398212.
$$

The second comment is that once we have exact closed-form expressions for the black hole degeneracies we can ask ourselves about their asymptotic limit and hence extract conclusions for macroscopic black holes. It is very important to realize that without such exact and closed-form expressions the problem of extracting all the relevant information in the macroscopic limit is very hard and some important features may actually be missed if coarse and difficult-to-control approximations are used.
An important feature of the black hole degeneracy spectrum that one would wish to recover in the macroscopic limit is the effective equi-spaced area spectrum found in \cite{val1}. In our opinion if such behavior is present it would be very strong evidence that LQG provides an accurate description of quantum gravity with the right semiclassical limit. We hope that the asymptotic analysis of the generating functions given above will uncover this type of behavior.

\begin{acknowledgments}
We want to thank I. Agull\'o, J. D\'{\i}az-Polo, and E. F. Borja for their comments, encouragement, and also for providing us with concrete numbers to cross-check the results presented here. This work is supported by the Spanish MEC grant FIS2005-05736-C03-02.

\end{acknowledgments}


\begin{thebibliography}{99}



\expandafter\ifx\csname
natexlab\endcsname\relax\def\natexlab#1{#1}\fi
\expandafter\ifx\csname bibnamefont\endcsname\relax
  \def\bibnamefont#1{#1}\fi
\expandafter\ifx\csname bibfnamefont\endcsname\relax
  \def\bibfnamefont#1{#1}\fi
\expandafter\ifx\csname citenamefont\endcsname\relax
  \def\citenamefont#1{#1}\fi
\expandafter\ifx\csname url\endcsname\relax
  \def\url#1{\texttt{#1}}\fi
\expandafter\ifx\csname urlprefix\endcsname\relax\def\urlprefix{URL
}\fi \providecommand{\bibinfo}[2]{#2}
\providecommand{\eprint}[2][]{\url{#2}}



\bibitem{abk}
  \bibinfo{author}{\bibfnamefont{A.} \bibnamefont{Ashtekar}},
\bibinfo{author}{\bibfnamefont{J.} \bibnamefont{Baez}},
\bibinfo{author}{\bibfnamefont{A.} \bibnamefont{Corichi}},
 \bibnamefont{and}
\bibinfo{author}{\bibfnamefont{K.} \bibnamefont{Krasnov}},
  \bibinfo{journal}{Phys. Rev. Lett.} \textbf{\bibinfo{volume}{80}},
  \bibinfo{pages}{904} (\bibinfo{year}{1998}).
\bibinfo{author}{\bibfnamefont{A.} \bibnamefont{Ashtekar}},
\bibinfo{author}{\bibfnamefont{J.} \bibnamefont{Baez}},
 \bibnamefont{and}
\bibinfo{author}{\bibfnamefont{K.} \bibnamefont{Krasnov}},
  \bibinfo{journal}{Adv. Theor. Math. Phys. } \textbf{\bibinfo{volume}{4}},
  \bibinfo{pages}{1} (\bibinfo{year}{2000}).


\bibitem{val1}
\bibinfo{author}{\bibfnamefont{A.} \bibnamefont{Corichi}},
\bibinfo{author}{\bibfnamefont{J.} \bibnamefont{Diaz-Polo}},
 \bibnamefont{and}
\bibinfo{author}{\bibfnamefont{E.} \bibnamefont{Fernandez-Borja}},
  \bibinfo{journal}{Phys. Rev. Lett.} \textbf{\bibinfo{volume}{98}},
  \bibinfo{pages}{181301} (\bibinfo{year}{2007}).
\bibinfo{author}{\bibfnamefont{A.} \bibnamefont{Corichi}},
\bibinfo{author}{\bibfnamefont{J.} \bibnamefont{Diaz-Polo}},
 \bibnamefont{and}
\bibinfo{author}{\bibfnamefont{E.} \bibnamefont{Fernandez-Borja}},
  \bibinfo{journal}{Class. Quant. Grav.} \textbf{\bibinfo{volume}{24}},
  \bibinfo{pages}{243} (\bibinfo{year}{2007}).

\bibitem{val2}
\bibinfo{author}{\bibfnamefont{I.} \bibnamefont{Agullo}},
\bibinfo{author}{\bibfnamefont{J.} \bibnamefont{Diaz-Polo}},
 \bibnamefont{and}
\bibinfo{author}{\bibfnamefont{E.} \bibnamefont{Fernandez-Borja}}.
\textit{Black hole state degeneracy in Loop Quantum Gravity}.
  \eprint{arXiv: 0802.3188}

\bibitem{BM}
\bibinfo{author}{\bibfnamefont{J. D.} \bibnamefont{Bekenstein}},
   \bibinfo{journal}{Lett. Nuovo Cimento} \textbf{\bibinfo{volume}{11}},
  \bibinfo{pages}{467} (\bibinfo{year}{1974}).
\bibinfo{author}{\bibfnamefont{V. F.} \bibnamefont{Mukhanov}},
   \bibinfo{journal}{JETP Letters} \textbf{\bibinfo{volume}{44}},
  \bibinfo{pages}{63} (\bibinfo{year}{1986}).
\bibinfo{author}{\bibfnamefont{J. D.} \bibnamefont{Bekenstein}}.
\textit{Classical Properties, Thermodynamics and Heuristic Quantization},
in Cosmology and Gravitation, M. Novello, ed. (Atlantisciences, France 2000), pp. 1-85.

\bibitem{prlnos}
  \bibinfo{author}{\bibfnamefont{I.} \bibnamefont{Agull\'o}},
\bibinfo{author}{\bibfnamefont{J.~F.} \bibnamefont{Barbero G.}},
\bibinfo{author}{\bibfnamefont{E.} \bibnamefont{Fernandez-Borja}},
\bibinfo{author}{\bibfnamefont{J.} \bibnamefont{D\'{\i}az-Polo}},
 \bibnamefont{and}
\bibinfo{author}{\bibfnamefont{E.~J.~S.} \bibnamefont{Villase\~nor}}.
\textit{Black hole state counting in Loop Quantum Gravity: A number theoretical approach}.
  \bibinfo{journal}{Phys. Rev. Lett. (to appear)},  \eprint{arXiv: 0802.4077}.


\bibitem{Burton}
\bibinfo{author}{\bibfnamefont{D. M.} \bibnamefont{Burton}}.
  \emph{\bibinfo{title}{Elementary Number Theory}}. \bibinfo{publisher}{McGraw-Hill, New York} \bibinfo{year}{(2002)}.

\bibitem{DM}
\bibinfo{author}{\bibfnamefont{R.~L.} \bibnamefont{Graham}},
\bibinfo{author}{\bibfnamefont{D.~E.} \bibnamefont{Knuth}},
\bibnamefont{and}
\bibinfo{author}{\bibfnamefont{O.} \bibnamefont{Patashnik}}.
  \emph{\bibinfo{title}{Concrete Mathematics: A Foundation for Computer Science}}. \bibinfo{publisher}{Addison-Wesley Professional; 2nd edition} \bibinfo{year}{(1994)}.
\bibinfo{author}{\bibfnamefont{K.~H.} \bibnamefont{Rosen}}.
  \emph{\bibinfo{title}{Discrete Mathematics and Its Applications}}.
  \bibinfo{publisher}{McGraw-Hill; 6th edition} \bibinfo{year}{(2006)}.

\bibitem{Kloster:2007cb}
\bibinfo{author}{\bibfnamefont{S.} \bibnamefont{Kloster}},
\bibinfo{author}{\bibfnamefont{J.} \bibnamefont{Brannlund}},
 \bibnamefont{and}
\bibinfo{author}{\bibfnamefont{A.} \bibnamefont{DeBenedictis}},
  \bibinfo{journal}{Class. Quant. Grav.} \textbf{\bibinfo{volume}{25}},
  \bibinfo{pages}{065008} (\bibinfo{year}{2008}).

\bibitem{DL}
\bibinfo{author}{\bibfnamefont{M.} \bibnamefont{Domagala}}
 \bibnamefont{and}
\bibinfo{author}{\bibfnamefont{J.} \bibnamefont{Lewandowski}},
  \bibinfo{journal}{Class. Quant. Grav.} \textbf{\bibinfo{volume}{21}},
  \bibinfo{pages}{5233} (\bibinfo{year}{2004}).
  \bibinfo{author}{\bibfnamefont{K.} \bibnamefont{Meissner}},
  \bibinfo{journal}{Class. Quant. Grav.} \textbf{\bibinfo{volume}{21}},
  \bibinfo{pages}{5245} (\bibinfo{year}{2004}).

\bibitem{GM}
\bibinfo{author}{\bibfnamefont{A.} \bibnamefont{Ghosh}}
 \bibnamefont{and}
\bibinfo{author}{\bibfnamefont{P.} \bibnamefont{Mitra}},
  \bibinfo{journal}{Phys. Lett. } \textbf{\bibinfo{volume}{B616}},
  \bibinfo{pages}{114} (\bibinfo{year}{2005}).






\end{thebibliography}
\end{document}